\documentclass[pre,aps,twocolumn,nofootinbib]{revtex4-1}
\usepackage[latin1]{inputenc}
\usepackage[english]{babel}
\usepackage{graphicx}
\usepackage{color}
\usepackage{amsmath}
\usepackage{amssymb}
\usepackage{multirow}
\usepackage{caption}
\usepackage[parfill]{parskip}    
\usepackage{graphicx}
\usepackage{amssymb}
\usepackage{epstopdf}
\usepackage{color}
\usepackage{graphicx}

\begin{document}

\title{\bf Exact solutions of Einstein equations for anisotropic  magnetic sources}

\author{E. Contreras}
\thanks{On leave from Universidad Central de Venezuela}
\email{ej.contreras@uniandes.edu.co}
\author{P. Bargue\~no}
\email{p.bargueno@uniandes.edu.co}
\affiliation{Departamento de F\'{\i}sica,
Universidad de los Andes, Apartado A\'ereo {\it 4976}, Bogot\'a, Distrito Capital, Colombia}

\author{G. Quintero Angulo}
\email{gquintero@fisica.uh.cu}
\affiliation{Facultad de F\'{\i}sica, Universidad de La Habana, \\San L\'azaro y L, Vedado, La Habana 10400, Cuba}
\author{A. Pérez Martínez}
\email{aurora@icimaf.cu}
\author{D. Alvear Terrero}
\email{dianaalvear@icimaf.cu}
\affiliation{Departamento de F\'{\i}sica Te\'orica, Instituto de Cibern\'etica, Matem\'atica y F\'isica (ICIMAF), Calle E esq 15 No. 309, Vedado, La Habana 10400, Cuba.}

\begin{abstract}
In this work, we have obtained exact solutions of Einstein equations for static and axially symmetric magnetized matter, 
specifically in plane-symmetric and almost-plane symmetric cases. Although these solutions impose constraints on the components 
of the energy-momentum tensor, some physically interesting situations, like the magnetized vacuum, might be described. 
Plane-symmetric solutions in presence of a non-vanishing cosmological constant have remarkable features. In particular, the 
system can be driven continuously to the isotropic case by an appropriate tuning of the cosmological constant, sweeping the 
magnetic field from weak to strong magnetic field regimes. The role of the cosmological constant in magnetic collapse is discussed 
with an emphasis in the description of jets from compact objects. For illustrative purposes, specific calculations for a 
magnetized electron gas have been developed.

\end{abstract}

\maketitle


\section{Introduction}

As pointed out in \cite{Kramers},  in the first four decades of research on general relativity the majority of exact solutions were obtained by solving the field equations under the assumption of spherical symmetry. The exterior and interior Schwarzschild solutions and the Friedmann model of relativistic cosmology are well-known examples. From the work of Taub in the mid nineties \cite{Taub1951},
the interest on plane--symmetric solutions (admitting a three parameter group of motions) has been raising continuously.
Specifically, plane--symmetric static perfect fluid solutions were reported during decades
\cite{Taub1956,Horsky1975,Teixeira1977,Bronnikov1979,Hojman1984,Davidson1987,Davidson1989}
for certain prescribed equations of state for isotropic fluids, some of them found as subcases of static cylindrically symmetric solutions studied for perfect fluids by Bronnikov \cite{Bronnikov1979}.
More recently, a detailed analysis of the general exact solution corresponding to a static and plane--symmetric distribution of matter with density proportional to pressure has been published \cite{Gamboa2008}. Note that, for the isotropic case, the energy--momentum tensor (EMT) can be written as $T^{\mu}_{\nu}=\textrm{diag}(-\varrho,P,P,P)$.

In spite of the interest for the isotropic case, the study of exact plane--symmetric solutions for anisotropic fluids has received considerably less attention.
Based in Ref.~\cite{Letelier1980}, solutions corresponding to an anisotropic and irrotational fluid described by two perfect--fluid components each one obeying the equation of state $P=\varrho$ \cite{Letelier1980bis,Krori1984} were reported.
Note that the kind of anisotropies considered in Refs. \cite{Letelier1980,Letelier1980bis,Krori1984} corresponds to an EMT of the form $T^{\mu}_{\nu}=\textrm{diag}(-\varrho,P_{x},P,P)$.

More general anisotropies were taken into account from the work of Vaidya and Som \cite{Vaidya1982}, where static plane--symmetric conformally coupled scalar fields with a traceless fluid--like EMT were reported. The effect of anisotropic stresses was also considered for the Einstein--Maxwell system outside a massive, electrically charged plane of infinite extension (including a nonvanishing cosmological constant)
\cite{Amundsen1983} with an EMT calculated in \cite{McVittie1929} of the form
$T^{\mu}_{\nu}=\textrm{diag}(-\varrho,\varrho,\varrho,-\varrho)$.
Following \cite{Vaidya1982}, a Casimir type EMT with $T^{\mu}_{\nu}=\textrm{diag}(-\varrho,3\varrho,-\varrho,-\varrho)$
was discussed in \cite{Gron1991}. In the context of stationary axisymmetric cases, solutions belonging to Carter's family 
$\left[A\right]$ representing an anisotropic fluid configuration have been reported in \cite{Papakostas2001} for the case $\varrho + P_{y} = 2 (P_{z}-P_{x})$. Recently, Gomes
\cite{Gomes2015} has studied the local form of static plane--symmetric spacetimes in the presence of matter when the pressures are linearly related to the energy density.

Another type of anisotropy appears when there is a constant magnetic field acting on a quantum gas. In this case, the spherical symmetry is broken and the EMT becomes anisotropic, with a splitting of the pressure in the direction parallel and transverse to the magnetic field, which means that there is also an anisotropy in the equations of state (EoS) of the star's fluid. Quantum gases are crucial in the description of astronomical objects, in particular for compact objects, whose magnetic fields may reach large values. Thus, a model to describe the structure of such magnetized compact object would need to go beyond spherical symmetry because the well-known Tolman-Oppenheimer-Volkoff equation (TOV) is not valid anymore.

The first attempts to describe a deformed compact object with cylindrical geometry were done numerically in 
Refs.~\cite{Paret:2015RAA,1674-4527-15-7-975} inspired in ideas presented in Ref.~\cite{Trendafilova2011EJPh}. 
Recently, in \cite{1742-6596-845-1-012005}, numerical solutions for deformed neutron stars were obtained. However, the approximation used in these models made impossible to determine the total mass of the stars. Instead, they obtained some insight related to the deformation associated to the anisotropic EoS.

Even more, other phenomena like pulsar jets might be connected to the anisotropies produced by the magnetic field \cite{HugoElizabethAN}. Jets are large prolate objects, sometimes with a length of millions of light years, that emerge from different astrophysical 
bodies like young stellar or compact objects. Although depending on their origin they have different size, scale and velocities, 
their structures are similar. The mechanism of formation of jets is still under debate, nevertheless the consensus is that the magnetic field plays an important role in their origin and propagation \cite{deGouveiaDalPino:2005xn}-\cite{deGouveiaDalPino:2004jy}.

Therefore, given the relevance of the anisotropies present in the magnetic field case, in this work we present exact solutions of Einstein equations, using non spherical symmetry, which could serve as a complementary step to the numerical computations towards the understanding of magnetized compact objects, including the description of astrophysical jets.

The manuscript is organized as follows. In section \ref{Sec2} the anisotropic EMT and EoS for a magnetized quantum system are 
presented and the quantum magnetic collapse is discussed. In section \ref{Sec3}, Weyl-type solutions are presented 
while section \ref{Sec4} is dedicated to the study of almost-plane-symmetric spacetimes. Section \ref{Sec5} is devoted to 
plane-symmetric solutions. Three cases are discussed, with emphasis in the role of the cosmological constant as a
parameter controlling both the anisotropy and the weak--strong field regime transition. 
In section \ref{Sec6} we conclude and make some final remarks.

\section{Anisotropic EMT and EoS for a magnetized quantum system}\label{Sec2}

In this section, we summarize the main features related to the EMT and the EoS for a magnetized quantum gas.
From the thermodynamical potential one gets an EMT given by \cite{PerezRojas:2006dq}

\begin{equation}\label{emtensor}
T^i_j=\frac{\partial\Omega}{\partial a_{i,\lambda}}a_{j,\lambda}-\Omega\delta_{j}^i,\quad\quad T_0^0=-\varrho,
\end{equation}
where $a_{i}$ denotes the boson or fermion fields, $\Omega$ stands for the thermodynamical potential over volume 
and $\varrho$ corresponds to minus the internal energy density.
Considering a thermodynamical potential that depends on an external electromagnetic field described by the tensor $F_k^j$, 
Eq.(\ref{emtensor}) leads to pressure terms of the form
\begin{equation}
\textit{T}^i_j=-\Omega-F_k^i\left( \frac{\partial\Omega}{\partial F_k^j}\right),\quad  i=j.
\end{equation}

Hence, the energy-momentum tensor for a fermion/boson gas at finite temperature under the influence of a constant magnetic field $\textbf{B}$ in the $z$ direction reads \cite{2000PhRvL..84.5261C}
	\begin{equation}\label{emtensorB}
	T^{\mu}_{\nu}=\text{diag} (-\varrho,P_{\perp},P_{\perp}, P_{\parallel}),
	\end{equation}
	where the magnitudes
\\
	\begin{subequations} \label{EoSEB}
		\begin{eqnarray}
		\varrho &=& \Omega + \mu N, \label{EoS1}\\
		P_{\perp} &=& -\Omega - B\mathcal{M} ,\label{EoS3} \\
		P_{\parallel} &=& - \Omega, \label{EoS2}
		\end{eqnarray}
	\end{subequations}
\\
correspond, respectively, to the energy density and to the perpendicular and parallel pressures of the magnetized system in the zero temperature limit, which is valid for quantum gases in astrophysical scenarios. Here, $\mu$ is the chemical potential, $N=-\partial\Omega/\partial\mu$ the particle density and $\mathcal{M}=-\partial\Omega/\partial B$ the magnetization.	

From Eqs. (\ref{EoSEB}) we can see explicitly the breaking of spherical symmetry. Furthermore, since $\Omega$ depends on the particle density and on the magnetic
field, Eq. (\ref{EoS3}) might be equal to zero, which means that for some values of $B$ and $N$, $P_{\perp}=0$ while $P_{\parallel} \neq 0$ and  $\varrho \neq 0$. In such a case it is said that the system undergoes a magnetic collapse \cite{2000PhRvL..84.5261C}. The possibility of having a zero perpendicular pressure for magnetized gases in astrophysical conditions has been shown by the explicit use of the equations of state in \cite{2000PhRvL..84.5261C,Felipe:2002wt,Aurora2003EPJC}.

Since a collapsed gas can be freely pushed towards the magnetic field direction while it exerts a non vanishing positive parallel pressure, this kind of instability has been connected to the astrophysical jet production \cite{HugoElizabethAN}. However, as
occurs when describing magnetized compacts objects, the general relativistic study of jets is incompatible with spherical symmetry. In this regard, we devote the next sections to look for non spherical solutions of Einstein equations that allow a general 
relativistic description of jets and magnetized compacts objects within an appropriate symmetry.

\section{Weyl--type solutions}\label{Sec3}

It is well known \cite{Kramers} that static and axially symmetric line elements can be parameterized as

\begin{eqnarray}
ds^2=-\chi^{2}dt^{2}+\alpha^{2}((dx^{1})^{2}+(dx^{2})^{2})+\beta^{2}d\phi^{2},
\end{eqnarray}

where $\alpha=\alpha(x^{1},x^{2})$, $\beta=\beta(x^{1},x^{2})$ and $\chi=\chi(x^{1},x^{2})$. The components of the Ricci tensor
satisfy

\begin{eqnarray}
R^{t}_{t}+R^{\phi}_{\phi}=\frac{\Delta(\beta\chi) }{\alpha^{2}\beta\chi},
\end{eqnarray}

being $\Delta$ the ``planar" Laplacian, $\Delta f=\frac{\partial f}{\partial x^{1}}+\frac{\partial f}{\partial x^{2}}$.

Let us assume a material content such that

\begin{equation}\label{tmunu}
 T^{t}_{t}+T^{\phi}_{\phi}-T^{\mu}_{\mu}=0.
\end{equation}

In this case we can choose Weyl's gauge by changing from $(x^{1},x^{2})$ to $(\rho,z)$ where $\rho=\beta \chi $ is harmonic and
$z$ is its conjugate. The line element can be written as

\begin{eqnarray}
ds^{2}=-e^{2\lambda}dt^{2} + e^{2\nu-2\lambda}(d\rho^{2}+dz^{2})+\rho^{2}e^{-2\lambda}d\phi^{2},
\end{eqnarray}
\\
where $\alpha=e^{\nu-\lambda}$, $\xi=e^{\lambda}$ and $\lambda$ and $\nu$ are both functions of the radial coordinate, $\rho$.

For simplicity, let us consider at this point that both $\nu$ and $\lambda$ are functions of $\rho$. 
If both $\nu$ and $\lambda$ are functions only of $z$, the trivial solution is obtained by requiring Weyl's gauge. Even more, although full dependence on $\rho$ and $z$ would be desirable in order to try to compute the total mass of the deformed object, the system of equations can not be exactly solved.

For a magnetized compact object, Weyl's gauge implies
\begin{equation}
P_{\perp}=0,
\end{equation}
which is of interest because it can naturally describe the transversal magnetic collapse of the object, as previously commented.

Moreover, Einstein equations lead to

\begin{subequations}
\begin{eqnarray}
&&e^{2 \lambda-2 \nu } \left(\nu ''-2 \lambda ''+\lambda '^2-2
\frac{\lambda '}{\rho}\right)=
-\varrho \\
&&\frac{\nu '}{\rho }-\lambda '^2=0 \\
&& \lambda '^2-\frac{\nu '}{\rho }=0 \\
&& e^{2 \lambda -2 \nu } \left(\lambda '^2+\nu ''\right)= P_{\parallel}.
\end{eqnarray}
\end{subequations}
\\
From these equations we can obtain formal relations for $\lambda$ and $\nu$ in terms of $\varrho$ and $P_{\parallel}$ given by
\\
\begin{subequations}
\begin{eqnarray}
\nu&=&\int_1^{\rho } \frac{P_{\parallel}(u)^2}{u (\varrho(u)+P_{\parallel}(u))^2} \, du+C_{1}\\
\lambda &=&\int_1^{\rho } \frac{P_{\parallel}(u)}{u (\varrho(u)+P_{\parallel}(u))} \, du+C_{2}
\end{eqnarray}
\end{subequations}
\\
together with the consistency constraint
\\
\begin{eqnarray}\label{cons}
\varrho(\rho ) P_{\parallel}'(\rho )-P_{\parallel}(\rho ) \varrho'(\rho )
=8\pi\frac{\rho  (\varrho(\rho )+P_{\parallel}(\rho ))^3}{e^{2 \lambda (\rho )-2 \nu (\rho )}}.
\end{eqnarray}
\\
Therefore, the problem will be formally solved at this point if an equation of state such that $\varrho$ and $P_{\parallel}$
satisfy Eq. (\ref{cons}) is provided. As the problem here is twofold (first find the equation of state and then
look for its
physical interest), and none of them are easy to attack, in the next sections we will tackle it by employing a different symmetry.

\section{Almost--plane--symmetric solutions} \label{Sec4}

Looking for exact solutions we impose the following requirements. First, we demand \mbox{
$T^{\mu}_{\nu}=\textrm{diag}(-\varrho,P_{1},P_{2},P_{\parallel})$} automatically.
This means that no crossed terms in the Einstein tensor are allowed, although we could eliminate them by introducing new constraints between the metric functions. Second, $P_{1}=P_{2}=P_{\perp}$ should hold automatically, which means that no additional terms are allowed as in the previous requirement, even we could eliminate them by introducing new constraints between the metric functions. Third, only static metrics will be considered and finally, Einstein field equations should be exactly soluble.

Given these assumptions, the line element must be either almost--plane--symmetric
\footnote{Almost--plane--symmetric solutions were first considered in Ref.
\cite{Brady1998} as a toy model to describe perturbations around the Cauchy horizon of a black hole without assuming any symmetry of the perturbed solution.} and
static

\begin{eqnarray}
ds^2= -g(z)(dt^2-dz^2)+f(x,y)(dx^2+dy^2)
\end{eqnarray}
\\
or plane--symmetric and static (see Sec.~\ref{Sec5}).



For almost--plane--symmetric solutions, Einstein equations read:

\begin{subequations}
\begin{eqnarray}
&&-\frac{(\partial_{y}f)^2+(\partial_{x}f)^2-f\left(\partial_{yy}f+\partial_{xx}f\right)}{2 f^3}=-\varrho \label{almostpsm} \\
&&-\frac{g'^2-g g''}{2 g^3}=P_{\perp}\label{almostpsm1},
\end{eqnarray}
\end{subequations}
\\
where $-\varrho =P_{\parallel}$ and the prime denotes derivative with respect to the $z$--coordinate. Although, in principle, $\varrho$, $P_{\parallel}$ and $P_{\perp}$ can be taken as general functions, the equations are not exactly soluble unless the energy density and the pressures are taken as constant quantities. More precisely, assuming $\varrho=\varrho_{0}$, $P_{\parallel}=P_{\parallel0}$
and $P_{\perp}=P_{\perp 0}$, the solution for Eqs. (\ref{almostpsm}) and (\ref{almostpsm1}) reads
\\
\begin{subequations}
\begin{eqnarray}
f(x,y)&=&\frac{(c_{3}^2+c_{4}^2)}{\varrho_{0}} \cosh^{-2}(c_{3}x+c_{4}y+c_{5})\label{fxy} \\
g(z)&=&-\frac{c_{1}}{4 P_{\perp 0}} \cosh^{-2}\bigg(\frac{\sqrt{c_{1}}}{2}(z+c_{2})\bigg),
\end{eqnarray}
\end{subequations}
\\

where $c_{1}$, $c_{2}$, $c_{3}$, $c_{4}$ and $c_{5}$ are integration constants. It is worth noticing that
the solution is constrained by $\varrho_{0}+P_{\parallel 0}=0$.

An alternative solution for $g(z)$ can be found taking \mbox{$P_{\perp}=0$} in Eq. (\ref{almostpsm1}), from where
\begin{equation}
g(z) = A e^{B z}.\label{otra}
\end{equation}
Note that, in this case,  $f(x,y)$ is given again by Eq. (\ref{fxy}) and we have $\varrho_{0}+P_{\parallel 0}=0$
as in the previous case.

The solutions Eqs. (\ref{fxy})-(\ref{otra}) are valid if $\varrho_{0}+P_{\parallel 0}=0$. Combining this condition with Eqs.~(\ref{EoSEB}), we get \mbox{$\mu N =0$}, i.e., there is no matter in the system. Therefore, it follows that
$\varrho = -P_{\parallel } = \Omega$. This is exactly what is obtained when describing a magnetized vacuum \cite{PerezRojas:2006dq}. Then, the analytic solution for the almost--plane--symmetric case presented in this section could be appropriate to study a vacuum under the action of an external magnetic field and the astrophysical and cosmological implications related to it.

\section{Plane--symmetric solutions} \label{Sec5}

To tackle plane--symmetric solutions, we will consider the same assumptions used previously in the case of almost--plane--symmetric geometry, and study some cases for different metric parametrizations.

\subsection{First case}

Considering the line element given  by\\
\begin{eqnarray} \label{metricVA}
\label{plane1}
ds^2= -g(z)(dt^2-dz^2)+f(z)(dx^2+dy^2),
\end{eqnarray}
\\
Einstein equations read
\\
\begin{subequations}
\begin{eqnarray}
-\frac{2 f f'g'+g \left(f'^2-4 f f''\right)}{4 f^2 g^2}&=&\varrho
\label{fgro} \\
-\frac{2 f^2 g'^2+g^2 \left(f'^2-2 f f''\right)-2 f^2 g g''}{4 f^2 g^3}&=&P_{\perp}
\label{fgp}\\
\frac{f'\left(g f'+2 f g'\right)}{4 f^2 g^2}&=&P_{\parallel}
\label{fgppara}
\end{eqnarray}
\end{subequations}
\\
We note that, in general, the above equations are not exactly integrable unless we take $g(z)=1$. 
In this particular case, if we consider a constant perpendicular pressure $P_{\perp}=P_{\perp0}$, 
with  $P_{\parallel} = P_{\parallel} (z)$ and $\varrho = \varrho (z)$, the solutions would be useful in a physical scenario 
only if the energy-momentum conservation condition is satisfied. For the metric (\ref{metricVA}) and the EMT (\ref{emtensorB}), 
this condition reads
\begin{equation}
	\frac{f'(P_{\parallel}-P_{\perp}) + f P'_{\parallel}}{f} = 0.
\end{equation}
After these considerations, we get meaningful solutions for $f(z)$, $\varrho (z)$ and $P_{\parallel}(z)$ after 
$P_{\perp}=0$ in  Eq.~(\ref{fgp}) is assumed. The corresponding solutions read
\\
\begin{eqnarray}
\label{jetplane}
f(z) &=& \frac{(c_{1}z+2c_{2})^2}{4 c_{2}}, \nonumber \\
\varrho (z) &=&-P_{\parallel} (z)= -\frac{1}{(2+z)^2},
\end{eqnarray}
\\
where energy-momentum conservation holds if $c_2=c_1$ and $P_{\parallel} + \varrho =0$ is satisfied. 
Since the latter constraint is exactly the condition discussed in the previous section, this solution might be useful 
also to describe a magnetized vacuum.

\subsection{Second case}

Considering the metric parametrized as
\begin{equation}
ds^2= -f(z)^{-1}dt^2+f(z)(dx^2+dy^2)+g(z)dz^2,
\end{equation}
we were able to obtain exact solutions of Einstein equations with
$\varrho=3 A_{0}$, $P_{\parallel}=-P_{\perp}=A_{0}$, where $A_{0}$ is certain constant and $g(z)$ is given by
\begin{equation}
g(z)=-\frac{f'(z)^2}{4 A_{0} f(z)^2},
\end{equation}
with $f(z)$ an arbitrary function.

The conditions $\varrho=3 A_{0}$, $P_{\parallel}=A_{0}$, $P_{\perp}=-A_{0}$ are equivalent to  $P_{\parallel}=\varrho/3$ and
$P_{\perp}=-\varrho/3$. Depending on the matter that composes the system, these equalities might be satisfied or not for certain
values of the particle density and magnetic field. 
It is worth noticing that $P_{\parallel}=\varrho/3$ is exactly the equation of state for a ultra-relativistic gas of non 
interacting particles. Even more, the condition holds true if the signs of $P_{\perp}$ and $P_{\parallel}$ are opposite. 
Therefore, the present solution might only describe an already collapsed gas. Finally, 
we have explored the phase space of an electron 
gas and we have found that, for magnetic fields between $10^9$ and $10^{13}$G, and for particles densities between 
$10^{30}$ and $10^{45}$cm$^{-3}$, the previous conditions are not satisfied. However, the applicability of the present solution 
for a different system can not be ruled out at this point.

\subsection{Third case}

In this section we will employ a very useful and general parametrization for plane--symmetric geometries as follows \cite{Dolgov1989}
\begin{equation}
ds^2= -f(z)dt^2+g(z)(dx^2+dy^2)+dz^2.
\end{equation}

By introducing the changes \mbox{$f(z)=p(z)^{2}q(z)^{-2/3}$}, \mbox{$g(z)=q(z)^{4/3}$}, Einstein equations
become

\begin{subequations}
\begin{eqnarray}
\frac{4 q''}{3 q}&=&-\varrho \\
\frac{p''}{p}+\frac{q''}{3 q}&=&P_{\perp} \\
\frac{4 p' q'}{3 p q}&=&P_{\parallel}
\end{eqnarray}
\end{subequations}

and we arrive to

\begin{subequations}
\begin{eqnarray}
q(z)&=&e^{-\frac{1}{2} \sqrt{3} \sqrt{A_1} z} \left(e^{\sqrt{3} \sqrt{A_1} z} c_1+c_2 \right) \\
p(z)&=&e^{-\frac{\sqrt{3} A_2 z}{2 \sqrt{A_1}}} \left(c_2-c_1 e^{\sqrt{3} \sqrt{A_1} z}\right)^{\frac{A_2}{A_1}} c_3,
\end{eqnarray}
\end{subequations}
\\
where $c_{1}$, $c_{2}$ and $c_3$ are constants of integration, \mbox{$\varrho= -A_{1}$} and $P_{\parallel}=A_{2}$. Even more, in this case, the EMT is given by

\begin{equation}\label{Tmunu}
T^{\mu}_{\nu}=\mathrm{diag}(A_{1},\frac{A_{1}^2+3 A_{2}^2 }{4 A_{1}},
\frac{A_{1}^2+3 A_{2}^2}{4 A_{1}},A_{2}).
\end{equation}
Note that in this case all the components of the EMT are constant and
not zero. As in previous cases, this specific form imposes a relation between the pressures and the energy density that 
restricts the external parameters of the system. However, this restriction can be relaxed with the inclusion of a cosmological constant, $\Lambda$, in the geometric sector of the equations.

We remind the reader that the cosmological constant has been introduced in other studies related to observable masses and radii
of compact objects because depending on it, the properties of these objects change
\cite{Mak2000,Balaguera2005,Balaguera2006,Zarro2009,Zarro2009bis,Stuchlik2012,Hossein2012,Kalam2012,Shojai2014,Wright2015,Nayak2015,ZubairiThesis,Zubairi2015,Bordbar2016}.
Even more, the inclusion of $\Lambda$ could  explain some processes that occurred in the early stage of the Universe, implying changes in the evolution of the first stars and, consequently, in the signals that are received
nowadays \cite{ZubairiThesis}.

With the addition of the cosmological constant, Einstein equations read
\begin{subequations}
\begin{eqnarray}
\frac{4 q''}{3 q}+ \Lambda&=&-\varrho \\
\frac{p''}{p}+\frac{q''}{3 q}- \Lambda&=&P_{\perp} \\
\frac{4 p' q'}{3 p q}- \Lambda&=&P_{\parallel},
\end{eqnarray}
\end{subequations}
\\
whose solutions are given by
\\
\begin{subequations}
\begin{eqnarray}
\label{cosmosol}
q(z)&=&e^{-\frac{1}{2} \sqrt{3} z \sqrt{A_{1}-\Lambda }} \left(e^{\sqrt{3} z \sqrt{A_{1}-\Lambda }} c_{1}+c_{2}\right)
 \\
p(z)&=&e^{\frac{(-A_{2}+\Lambda ) \left(3 z \sqrt{A_{1}-\Lambda}-2 \sqrt{3} \mathrm{log} \left[c_{2}-c_{1} e^{\sqrt{3}
z \sqrt{A_{1}-\Lambda }}\right]\right)}{2 \sqrt{3} (A_{1}-\Lambda )}} c_{3}, \nonumber
\end{eqnarray}
\end{subequations}
\\
where $c_{1}$, $c_{2}$ and $c_3$ are again certain constants of integration.
\\
With this solution, the EMT can be written as
\begin{widetext}
\begin{equation}\label{Tmunu1}
T^{\mu}_{\nu}=\mathrm{diag}(A_{1},\frac{A_{1}^2+3 A_{2}^2+2 A_{1} \Lambda -6 A_{2} \Lambda }{4 A_{1}-4 \Lambda },
\frac{A_{1}^2+3 A_{2}^2+2 A_{1} \Lambda -6 A_{2} \Lambda }{4 A_{1}-4 \Lambda },A_{2}).
\end{equation}
\end{widetext}

From Eq.~(\ref{Tmunu1}) we can identify

\begin{equation}
	P_{\perp}=-\frac{\rho^2 + 3P_{\parallel}^2  -2\rho\Lambda -6P_{\parallel }\Lambda }{4\rho+4\Lambda}.
\end{equation}
\\
Thus, as a function of the energy and the pressures, the cosmological constant reads
\begin{equation}\label{lambda}
	\Lambda=\frac{3P_{\parallel}^2+\rho^2+4 P_{\perp}\rho}{2(3P_{\parallel }+\rho-2P_{\perp})}.
\end{equation}

It is worth mentioning that by an appropriate tuning of $\Lambda$ in terms of $N$ and $B$, the collapsing solution could be reached.
This specific case for which $P_{\perp}=0$ can be obtained when

\begin{equation}\label{Lambdacollapse}
	\Lambda = \frac{3 P_{\parallel}+\varrho^2}{2 (3 P_{\parallel} + \varrho)}.
\end{equation}

Furthermore, we note that $\Lambda$ allows us to recover the isotropic case from the anisotropic one.
Setting $P_{\perp} = P_{\parallel}$ in Eq. (\ref{lambda}) we found that $\Lambda$ must have the value

\begin{equation}\label{Lambdaiso}
	\Lambda = (\varrho+3 P_{\parallel})/2,
\end{equation}
\noindent to account for the isotropy. This is a very important result that could be of particular relevance for
modeling anisotropic compact objects, since it means that not only $\Lambda$ can be used to control the anisotropy,
but also that the system could be driven continuously to the isotropic case.

To get some insight on the order of magnitude of the cosmological constant and to study its dependence on the external
parameters of a magnetized quantum gas, we have computed $\Lambda$ using the EoS for a gas of electrons in a magnetic field given by the equations  \cite{Felipe:2002wt}

\begin{eqnarray}\label{220}
  \varrho &=& \frac{m^2}{4\pi^2}\frac{B}{B_c} \sum_{l=0}^{l_{max}} g_{l}\! \left( \mu\,p_F +{\mathcal {E}_{l}}^2\ln\frac{ {\mu}+ {p_F}}{\mathcal {E}_{l}}\right),\\
  P_{\parallel}
  &=& \frac{m^2}{4\pi^2}\frac{B}{B_c}\sum_{l=0}^{l_{max}}g_{l}\!\left[ \mu\,p_F -{\mathcal {E}_{l}}^2\ln\!\left(\frac{\mu+ p_F}{\mathcal {E}_{l}}\right)\right]\!,\\
  P_{\perp} &=& \frac{m^4}{2\pi^2}\left(\frac{B}{B_c}\right)^{\!\!2}\,\sum_{l=0}^{l_{max}}g_l l\ln\left (\frac{\mu+p_F}{\mathcal{E}_l}\right),
\end{eqnarray}

 \noindent where $l$ are the Landau levels, $l_{max}= I[\frac{\mu^2-m^2}{2eB}]$ and $I[z]$ denotes the integer part of $z$. The Fermi momenta is ${p_F}=\sqrt{{\mu}^2-\mathcal{E}_{l}^2}$, with the rest energy given as $\mathcal{E}_{l}=\sqrt{2|eB|l+m^2}$, and the quantity $B_c = m^2/e \sim 4 \times 10^{13} G$ corresponds to the so-called critical magnetic field or Schwinger field.

\begin{figure}[!h]
	\centering
	\includegraphics[width=0.95\linewidth]{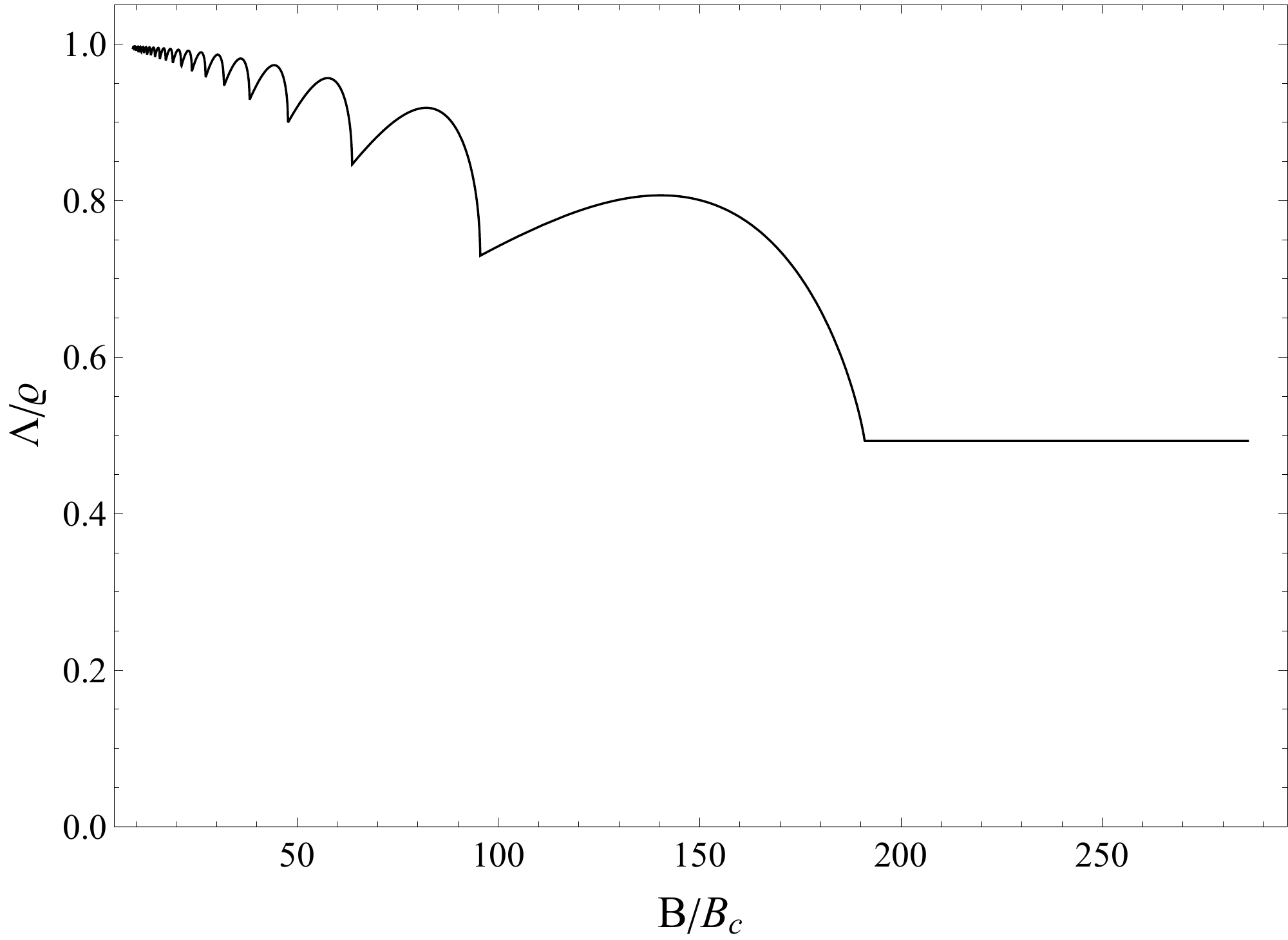}
	\caption{Behavior of the ratio $\Lambda/\rho$ as a function of $B$ for fixed $N$ ($\mu=19.6 m_e$). 
Note the Haas van Alphen oscillations due to the presence of Landau levels. Weak and strong magnetic fields regimes correspond
to $\Lambda/\rho \to 1$ and to $\Lambda/\rho \to 1/2$ in the left and right sides of the figure, respectively.}
    \label{fig1}
\end{figure}

\begin{figure}[!h]
	\centering
	\includegraphics[width=0.95\linewidth]{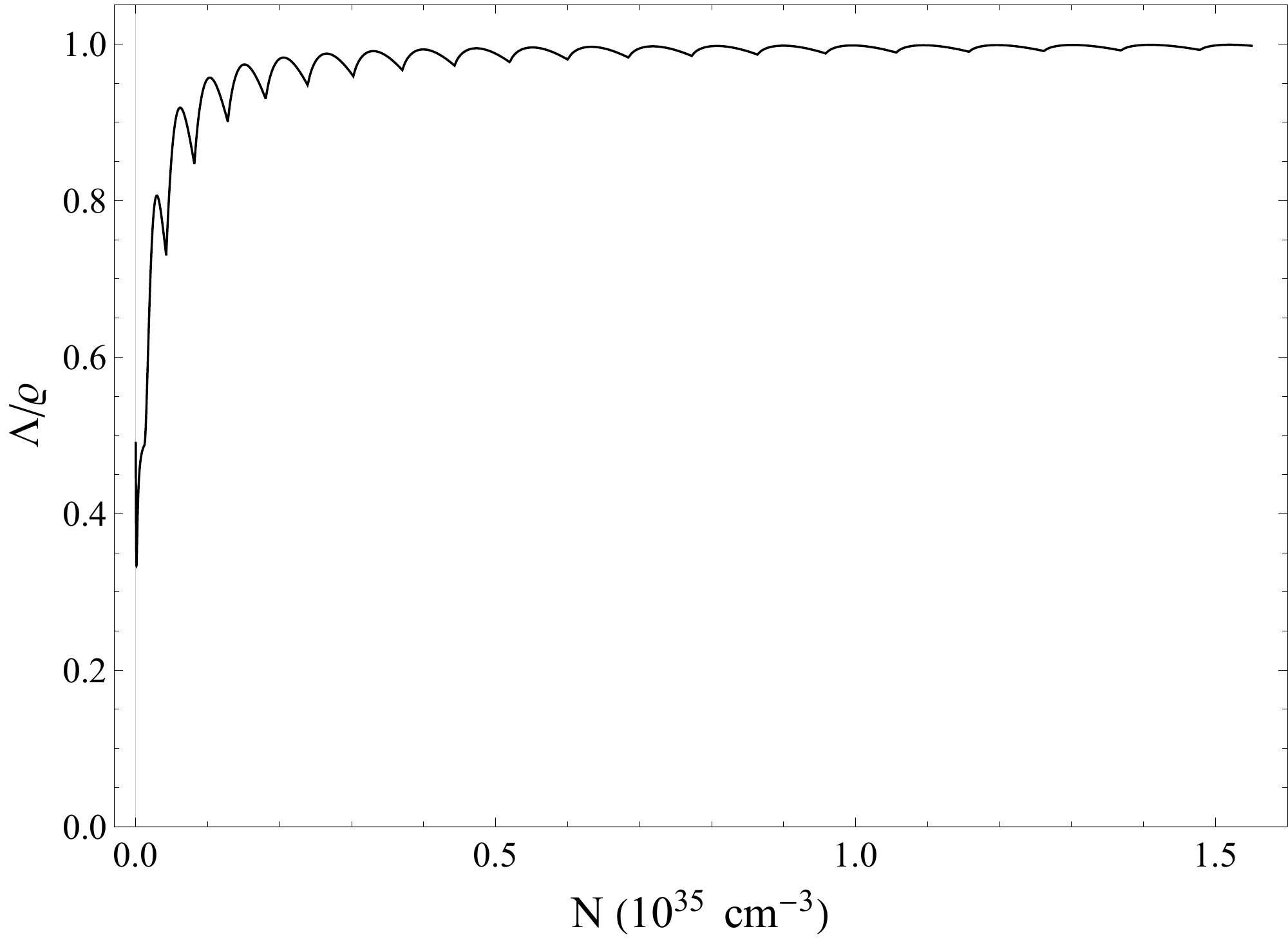}
	\caption{Behavior of the ratio $\Lambda/\rho$ versus $N$ for fixed $B = 100 ~B_c$. Note the Haas van Alphen oscillations. 
}
    \label{fig2}
\end{figure}

On one hand, in Fig. (\ref{fig1}) we plot the ratio $\Lambda/\varrho$ as a function of the magnetic field at a
given particle density. On the other hand, Fig. (\ref{fig2}) shows $\Lambda/\varrho$ depending on the particle density for a 
fixed value of the magnetic field. We note that the values of $N$ and $B$ that appear in both figures are standard for 
compact objects. In addition, both figures show the Haas van Alphen oscillations due to the presence of Landau levels.
Let us note that \mbox{$0 < \Lambda/\varrho < 1$}.


Fig. (\ref{fig1}) is especially illustrative because it encloses two extreme cases. For a strong
magnetic field all particles are accommodated
in the lowest Landau level and the gas has collapsed, $P_{\perp}=0$. In this limit, $\Lambda/\varrho = 1/2$. As the magnetic
field decreases, more particles are populating higher Landau levels. In this regime, the
Haas van Alphen oscillations appear and $P_{\perp} \neq 0$. If the magnetic field continues decreasing, the number of Landau 
levels increases while the spacing among them tends to form a continuum. When the magnetic fields is turned to zero, 
$\varrho \to 3 P_{\parallel}$ for the particles here discussed.
In consequence, $(\varrho+3 P_{\parallel})/2 \to \varrho$ and $\Lambda / \varrho \to 1$, as Fig. (\ref{fig1}) shows.
\\
Therefore, as can be read from the EoS of Eqs. (\ref{EoSEB}) and from Einstein equations, including a cosmological constant 
produces interesting effects not only in terms of anisotropies but also in terms of the weak--strong field transition.

\section{Final remarks and conclusions}\label{Sec6}

In this work we have obtained exact solutions of Einstein equations for static and axially symmetric cases
considering that this symmetry is inherited from magnetized matter.
Solutions are obtained for both plane--symmetric and almost--plane symmetric geometries.
These solutions impose some constraints between the energy density and the anisotropic pressures which make very restrictive the
situations where they are fulfilled. However, we have also studied the plane--symmetric case with the inclusion of the 
cosmological constant, $\Lambda$. It is worth noticing that not only the anisotropy can be ``controlled" by the cosmological 
constant, but the system could also be driven continuously to the isotropic case by an appropriate tuning of it. We have found the 
asymptotic behaviour for $\Lambda$ for both weak ($\Lambda/\rho\to 1$) and strong field ($\Lambda/\rho\to 1/2$) regimes, 
the latter corresponding to the case when the magnetic collapse  appears ($P_{\perp}=0$), which is of importance to describe jets from compact objects. These features have been illustrated for a magnetized electron gas.

\section{Acknowledgements}

The authors D. A. T, A. P. M and G. Q. A acknowledge to Hugo P\'erez the inspiration to tackle part of this work, and also 
thank the comments of E. Rodr\'{\i}guez Querts on this manuscript. E. C. and P. B. acknowledge 
the support from the Faculty of Science and
Vicerrector\'{\i}a de Investigaciones of Universidad de Los Andes, Bogot\'a, Colombia.

\end{document}